\documentclass[aps,prl,superscriptaddress,reprint,floatfix]{revtex4-1}
\usepackage[utf8]{inputenc}
\usepackage{amsmath}
\usepackage{amssymb}
\usepackage{graphicx}
\usepackage[colorlinks,urlcolor=blue]{hyperref}
\usepackage{url}
\usepackage{color,xcolor}
\usepackage{ulem}
\usepackage{float}
\usepackage{epstopdf}
\usepackage[none]{hyphenat}
\begin{document}

\title{Layer-dependent $s^{\pm}$ superconducting pairing and influence of incipient bands in La$_3$Ni$_2$O$_7$ thin films on real LaSrAlO$_4$ substrates}
\author{Yang Zhang}
\affiliation{Materials Science and Technology Division, Oak Ridge National Laboratory, Oak Ridge, Tennessee 37831, USA}
\author{Philip M. Dee}
\affiliation{Computational Sciences and Engineering Division, Oak Ridge National Laboratory, Oak Ridge, Tennessee 37831, USA}
\author{Ling-Fang Lin}
\affiliation{Department of Physics and Astronomy, University of Tennessee, Knoxville, Tennessee 37996, USA}
\author{Adriana Moreo}
\affiliation{Department of Physics and Astronomy, University of Tennessee, Knoxville, Tennessee 37996, USA}
\affiliation{Materials Science and Technology Division, Oak Ridge National Laboratory, Oak Ridge, Tennessee 37831, USA}
\author{Thomas A. Maier}
\affiliation{Computational Sciences and Engineering Division, Oak Ridge National Laboratory, Oak Ridge, Tennessee 37831, USA}
\author{Elbio Dagotto}
\affiliation{Department of Physics and Astronomy, University of Tennessee, Knoxville, Tennessee 37996, USA}
\affiliation{Materials Science and Technology Division, Oak Ridge National Laboratory, Oak Ridge, Tennessee 37831, USA}

\date{\today}

\begin{abstract}
Motivated by recent advances in ambient-pressure superconducting La$_3$Ni$_2$O$_7$ (LNO) thin films, we theoretically investigate these films by explicitly incorporating the realistic LaSrAlO$_4$ (LSAO) substrate. The Fermi-surface topology of the nickelate thin film closely resembles that of the high-pressure bulk phase, with similar $\alpha$ and $\beta$ sheets and a small hole-like $\gamma$ pocket originating from the surface bilayer (BL3). The bonding $d_{3z^2-r^2}$ state of the interfacial bilayer (BL1, adjacent to the LSAO substrate) lies only $\sim 5$ meV below the Fermi level, giving rise to an incipient $\gamma$ band, while for the middle bilayer (BL2) its energy is much lower ($\sim 95$ meV). Remarkably, no significant spontaneous charge transfer from the LSAO substrate to the LNO film is found. Instead, {\it symmetry-allowed charge disproportionation} occurs between the BLs, with hole doping in BL3 and electron doping in BL1 and BL2. Via the weak-coupling Random Phase Approximation, we find a strong $s^{\pm}$-wave superconducting pairing instability in BL3, while $d_{x^2-y^2}$ and $s^{\pm}$ are nearly degenerate in the other bilayers BL1 and BL2. Including the incipient $\gamma$ band via the Dynamical Cluster Approximation beyond weak coupling, robust $s^{\pm}$-wave pairing is observed in all three layers.
\end{abstract}

\maketitle

{\it Introduction--}
The discovery of high-pressure superconductivity in the Ruddlesden–Popper (RP) nickelate bilayer (BL) La$_3$Ni$_2$O$_7$ (LNO) with $T_c \sim 80$ K~\cite{Sun:arxiv} and trilayer (TL) La$_4$Ni$_3$O$_{10}$ with $T_c \sim 20–30$ K~\cite{Sakakibara:arxiv09,Li:cpl,Zhu:arxiv11,Huang:cpl-tri,Zhang:arxiv11}, has opened a new platform for exploring unconventional superconductivity~\cite{Zhang:arxiv-exp,Zhang:prb23,Wang:arxiv9,Dong:arxiv12,Yang:arxiv09,Wang:nature,Wang:jacs,LiuZhe:arxiv,Xie:SB,Chen:arxiv2024,Zhang:prb23-2,Sakakibara:prl24,Oh:prb24,Ryee:prl,Braz:arxiv25,Qu:prl,Lu:prl,Liu:arxiv2023,Zhang:prb24,Liu:arxiv,Yang:arxiv1,Liao:arxiv,Zhang:arxiv24,Zhang:nc24,Qin:arxiv,Lin:magnetism,Li:nature2026,Dong:arxiv25,Zhang:2025hp,Wang:cplreview,Wang:nsr2025,Puphal:nrp2025}. Subsequently, two hybrid-stacking RP nickelates were synthesized in bulk form: the 1313 phase with alternating single-layer (SL) and TL blocks~\cite{Chen:jacs1313,Puphal:arxiv12,Wang:ic,Abadi:arxiv24}, and the 1212 phase, composed of alternating SL-BL blocks~\cite{Li:prm24}. Similarly to BL and TL RP nickelates, superconductivity is absent in the bulk form at ambient pressure; however, superconductivity emerges under high pressure, with $T_c \sim 64$ K in the 1212 nickelate~\cite{Shi:arxiv25} and $T_c \sim 3.6$ K in the 1313 structure~\cite{Huang:arxiv25-1313}, further broadening the study of the RP nickelate family~\cite{Zhang:1313,LaBollita:1313,Lechermann:PRM2024,Chen:arxiv26,Zhang:2323,Zhang:arXiv2506-1212,Shao:nc26,Zhang:1212,Zhang:1313-film,Au-Yeung:NP,Xie:1212,Feng:1212,Zhangm:1313}.

Alongside rapid progress in high-pressure RP bulk nickelates, superconductivity has also been observed under ambient conditions in LNO thin films grown on the compressive-strain LaSrAlO$_4$ substrate (LSAO)~\cite{Ko:nature,Zhou:nature}. More recently, superconductivity was also reported in 1212 and 2323 nickelate films on the LSAO substrate without extra pressure~\cite{Nie:arxiv2025-apres}. These exciting discoveries mark an important advance in nickelate superconductivity and stimulated growing interest in film studies~\cite{Bhatt:arxiv2025,Wang:arxiv2502,Liu:nm25,Hao:arxiv25,Ji:arxiv25,Osada:cp25,Tarn:arxiv25,Shen:arxiv25,Li:nsr25,Wang:arxiv25,Zhang:prb26,Fan:STM,Huang:arxiv25-LPNO,Qiu:arxiv25,Cao:arxiv25,Bleys:arxiv25,Zhou:arxiv2025-60K,Liu:arxiv2026-dome,Sun:arxiv2025,Han:arxiv2026-film,Li:arxiv2025-film,Zhao:prb25,Yi:PRB2025,Geisler:arxiv25,Hua:arxiv2026,Yao:extra,Zhang:film-review,Zhang:exp-film-review,Hsu:nc26,Shi:AM26,Shao:prb25,Shao:arxiv25,Ryee:prl25,Ji:prb25-strong,Yue:NSR25,Chen:arxiv2026-film,Cao:arxiv2026-lao,Ushio:arxiv25,Osada:nm26,Chen:arxiv2026-lpno,Liang:arxiv2026-lpno,Osada:arxiv2026-pno,Wang:arxiv2026-lpco,Miao:arxiv2026-rp}. In addition, the realization of ambient pressure superconductivity in nickelate thin films allows the direct investigation of their electronic structures and superconducting state by using angle-resolved photoemission spectroscopy (ARPES). However, ARPES results remain controversial as to whether the small hole-like $\gamma$ pocket, originating from the bonding $d_{3z^2-r^2}$ orbital, crosses the Fermi level~\cite{Shen:arxiv25,Li:nsr25,Wang:arxiv25,Sun:arxiv2025}.

This controversy also persists in theoretical studies. Using a single BL LNO structure with in-plane lattice constants fixed to those of LSAO within two symmetry-equivalent Ni-oxide planes, several studies reported an $s^{\pm}$ pairing state associated with the emergence of the $\gamma$ pocket~\cite{Bleys:arxiv25,Shao:prb25,Cao:arxiv25,Shao:arxiv25,Yue:NSR25,Huang:arxiv25-LPNO} due to Fermi-surface nesting similar to the bulk system under high pressure. In contrast, Ref.~\cite{Zhang:prb26} found that the $\gamma$ pocket is absent when using a one-unit-cell structural model with fixed in-plane lattice constants, while upon hole doping, the $\gamma$ pocket emerges together with a $d$-wave superconducting instability. The symmetry equivalence between the top and bottom bilayers is broken in the one-unit-cell structural model due to the asymmetric environment created by the nickelate–vacuum interface (note that each LNO unit-cell involves two bilayers).

\begin{figure*}
\centering
\includegraphics[width=0.8\textwidth]{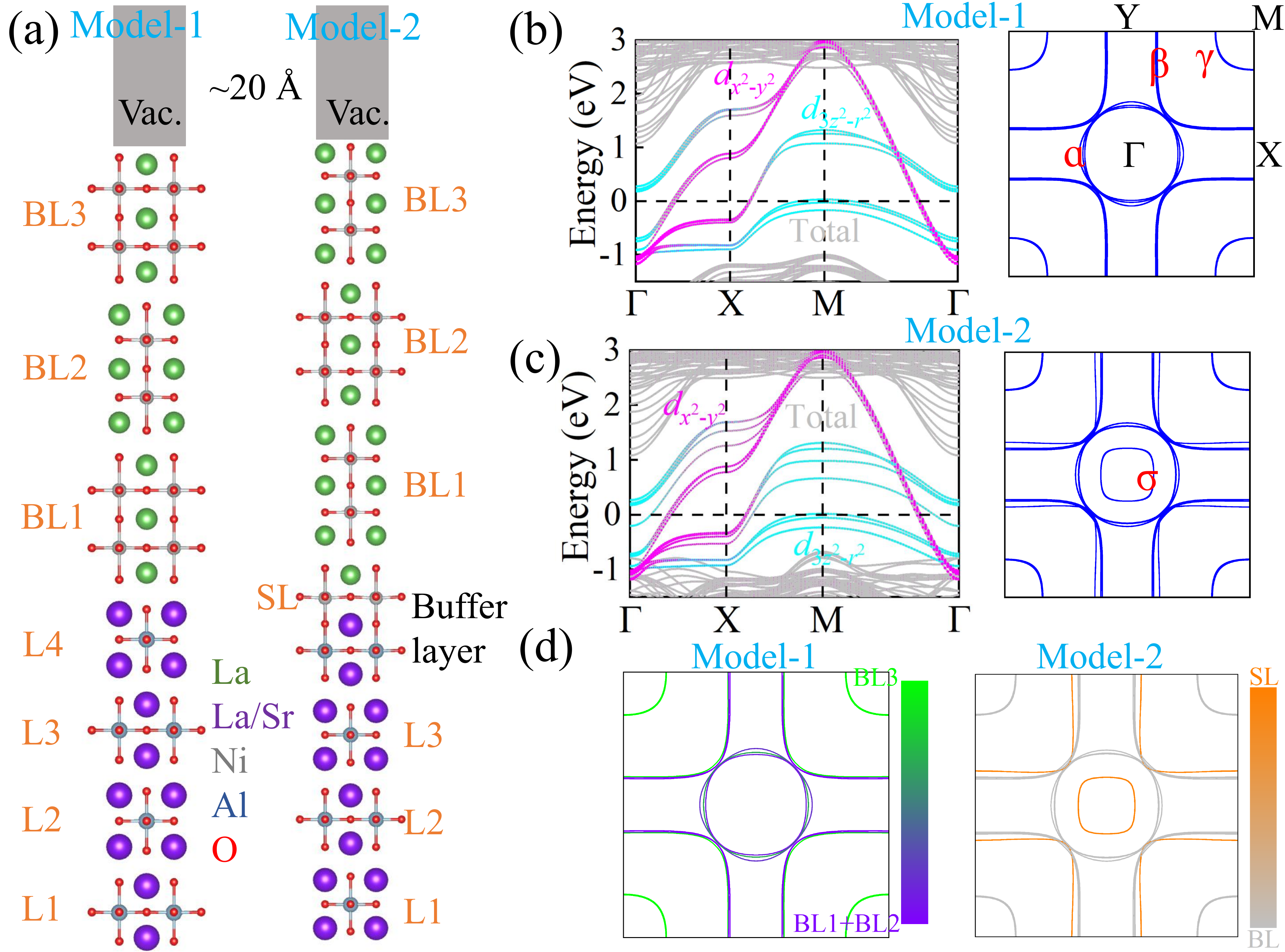}
\caption{(a) Schematic structural models studied here for BL nickelate thin films on the LSAO substrate (green = La, purple= La/Sr, gray = Ni, red = O), visualized using the VESTA code~\cite{Momma:vesta}. Model 2 has an additional buffer layer compared to Model 1.  (b-c) Projected band structures and their Fermi surfaces for the BL nickelate thin film on the LSAO substrate. We calculated the electronic structure by using density functional theory (DFT)~\cite{Kresse:Prb,Kresse:Prb96,Blochl:Prb,Perdew:Prl} plus Hubbard $U$ and Hund’s coupling $J$ within the Liechtenstein formulation~\cite{Liechtenstein:prb} (see more details in Supplementary Material~\cite{Supplemental}). Here, we use $U = 3.8$ eV and $J = 0.6$ eV from the constrained random-phase approximation for LNO~\cite{Christiansson:arxiv}. The high-symmetry points  of the Brillouin zone are $\Gamma$ = (0, 0, 0), X = (0, 0.5, 0), and M = (0.5, 0.5, 0). (d) Layer-resolved contributions of the two structural models to the Fermi surface states.}
\label{Structural-slab}
\end{figure*}

In a real thin film grown on LSAO, two distinct interfaces are present: the nickelate–vacuum interface at the top surface and the nickelate–LSAO interface at the bottom, while the interior BLs are connected by adjacent BLs. Without explicitly including the realistic substrate in theoretical calculations, important questions cannot be directly addressed, such as what is the origin of the hole-doping mechanism of the $\gamma$ pocket? In many publications, it was {\it assumed} that there was a spontaneous charge transfer from the substrate to the nickelate film. Is this correct? What is the nature of the pairing channel, and how does superconductivity, when LSAO is incorporated, link to results from previous studies that merely fixed the lattice constants to those of LSAO? Despite these important questions, systematic studies based on realistic thin-film geometries with the LSAO substrate remain rare~\cite{Geisler:arxiv25-1,Zhang:arxiv2610.03601}.

{\it Structural models--} To address the questions above, we first considered two slab models with the real LSAO substrate, as shown in Fig.~\ref{Structural-slab}(a): (1) Model-1 consists of four LSAO layers and three BL nickelate blocks; (2) Model-2 consists of four LSAO layers, one Ni (113) buffer layer, and three BL nickelate blocks. The reason for using Model-2 is that some experiments use a buffer layer~\cite{Ko:nature,Zhou:nature}. The in-plane lattice constants were fixed to those of the LSAO substrate, and a vacuum layer thicker than 20.0 \AA ~is introduced to simulate the thin-film slab geometry. Furthermore, the virtual crystal approximation~\cite{Bellaiche:Prb,Ramer:Prb} was employed to treat the mixed La and Sr sites in the LSAO substrate.

As shown in Fig.~\ref{Structural-slab}(b), for both Model-1 and Model-2, the two Ni-$e_g$ orbitals contribute to the low-energy electronic bands crossing the Fermi level, whereas the electronic states from the LSAO substrate are located at deeper energies. In addition, the Fermi-surface topology of Model-1 closely resembles that of the bulk phase under high pressure~\cite{Zhang:prb23}. Specifically, a small hole-like $\gamma$ pocket emerges around the M point for both cases, in addition to the usual $\alpha$ and $\beta$ sheets. By analyzing the layer-resolved contributions of the Fermi surface in Fig.~\ref{Structural-slab}(d), this hole pocket arises from the {\it surface BL}. The presence of the buffer layer introduces an additional electron-like $\delta$ pocket centered at the $\Gamma$ point and an additional hole-like Fermi surface sheet around the X and Y points, as illustrated in Fig.~\ref{Structural-slab}(d). However, the Fermi surface topology of the BL nickelate films remains essentially unchanged. This suggests that the buffer layer has little effect on the electronic structure of the nickelate thin films~\cite{Supplemental}. Therefore, we focus only on the structural Model-1 for the remainder of this work.

Crucially, {\it we do not observe spontaneous charge transfer from the LSAO substrate to the BL nickelate layers}, within our accuracy. However, a (weak) charge disproportionation is found among the inequivalent Ni BLs.   By using the original DFT Fermi level in the Wannier Hamiltonian, the average electronic densities of $e_g$ orbitals are 2.89, 3.07, and 3.02 for the BL3, BL2, and BL1 blocks, respectively, indicating hole doping at the surface BL3 and mild electron doping in the middle BL2 and interfacial BL1, compared to the stoichiometric LNO ($n = 3$). This asymmetry is similar to that reported for the infinite-layer nickelate on a SrTiO$_3$ substrate~\cite{Zhang:prb20}. Thus, the primary intuitive reason why BL3 hosts a $\gamma$ pocket is that its hole doping shifts the bonding $d_{3z^2-r^2}$ state upward so it crosses the Fermi level. Thus, the $\gamma$ pockets observed in some ARPES thin-film experiments likely originate at the top BL of the film, and not from a transfer from LSAO, as widely assumed before. This is also compatible with experiments that show that using LAO as a substrate (not LSAO) also leads to superconductivity~\cite{Tarn:arxiv25}, in spite of the absence of Sr$^{2+}$ in the latter.

{\it Tight-binding models--} 
To understand the low-energy physics of the LNO thin film, we constructed a four-band $e_g$-orbital tight-binding model for the Model-1 structure, extracting the crystal-field splittings and hopping integrals from maximally localized Wannier functions (MLWFs)~\cite{Mostofi:cpc}. The inter-BL hopping amplitudes are negligibly small and are therefore neglected; we then constructed three independent sub-BL models.

\begin{figure} 
\centering
\includegraphics[width=0.40\textwidth]{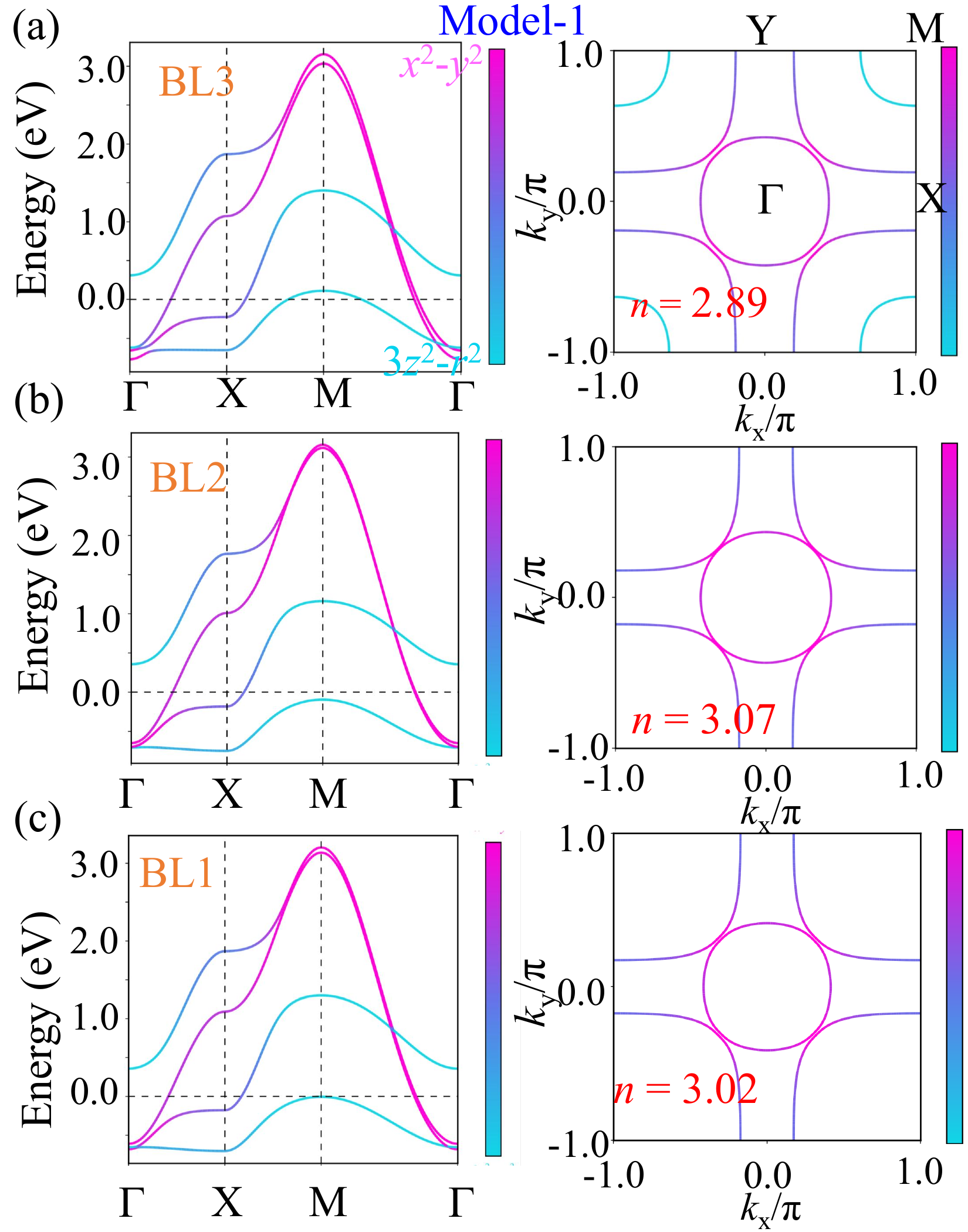}
\caption{Tight-binding band structures and Fermi surfaces for (a) BL3, (b) BL2, and (c) BL1 with its corresponding electronic densities, respectively. The hopping files of each BL and additional details for the tight-binding model used here can be found in the Supplementary Material~\cite{Supplemental}. }
\label{TB}
\end{figure}

As shown in Fig.~\ref{TB}, the tight-binding band structures and corresponding Fermi surfaces qualitatively agree with the DFT results discussed in the previous section. For all three BLs, the intralayer hopping between the $d_{x^2-y^2}$ orbitals remains nearly unchanged due to the epitaxial constraint imposed by the LSAO substrate. However, the interlayer hopping between the $d_{3z^2-r^2}$ orbitals substantially differs among the three BLs. The ratio $t^z_{\alpha\alpha}/t^{x/y}_{\beta\beta}$ ($\alpha=d_{3z^2-r^2}$ and $\beta=d_{x^2-y^2}$) is suppressed to $\sim$1.19 in the BL3, compared to that in the BL2 ($\sim 1.28$) and BL1 ($\sim 1.27$). Figure~\ref{TB}(a) shows that {\it the $\gamma$ pocket arises from the $d_{3z^2-r^2}$ bonding state in BL3}. In contrast to BL3, the bonding $d_{3z^2-r^2}$ states in BL2 and BL1 lie {\it below} the Fermi level (by $\sim 95$ and $\sim 5$ meV, respectively), and, therefore, the $\gamma$ pocket is absent for these bilayers, as shown in Figs.~\ref{TB}(b) and (c).

{\it RPA Results--} 
To explore the superconducting pairing instability in the LNO thin film on the LSAO substrate, we first performed a weak coupling analysis using multi-orbital Random Phase Approximation (RPA) calculations ~\cite{Kubo2007,Graser2009,Altmeyer2016,Romer2020} for each BL tight-binding model. In this formalism, the pairing strength $\lambda_\alpha$ and the gap structure $g_\alpha({\bf k})$ for channel $\alpha$ are obtained by solving an eigenvalue problem of the form
\begin{eqnarray}\label{eq:pp}
	\int_{FS} d{\bf k'} \, \Gamma({\bf k -  k'}) g_\alpha({\bf k'}) = \lambda_\alpha g_\alpha({\bf k})\,.
\end{eqnarray}
The momenta ${\bf k}$ and ${\bf k'}$ are restricted to the Fermi surface, and the singlet pairing interaction $\Gamma({\bf k- k'})$ is given by the irreducible particle-particle vertex. Within RPA, the dominant term entering $\Gamma({\bf k-k'})$ is the RPA spin susceptibility $\chi({\bf k-k'})$. Extra computational details for RPA are provided in the SM~\cite{Supplemental}.

\begin{figure}
\centering
\includegraphics[width=0.42\textwidth]{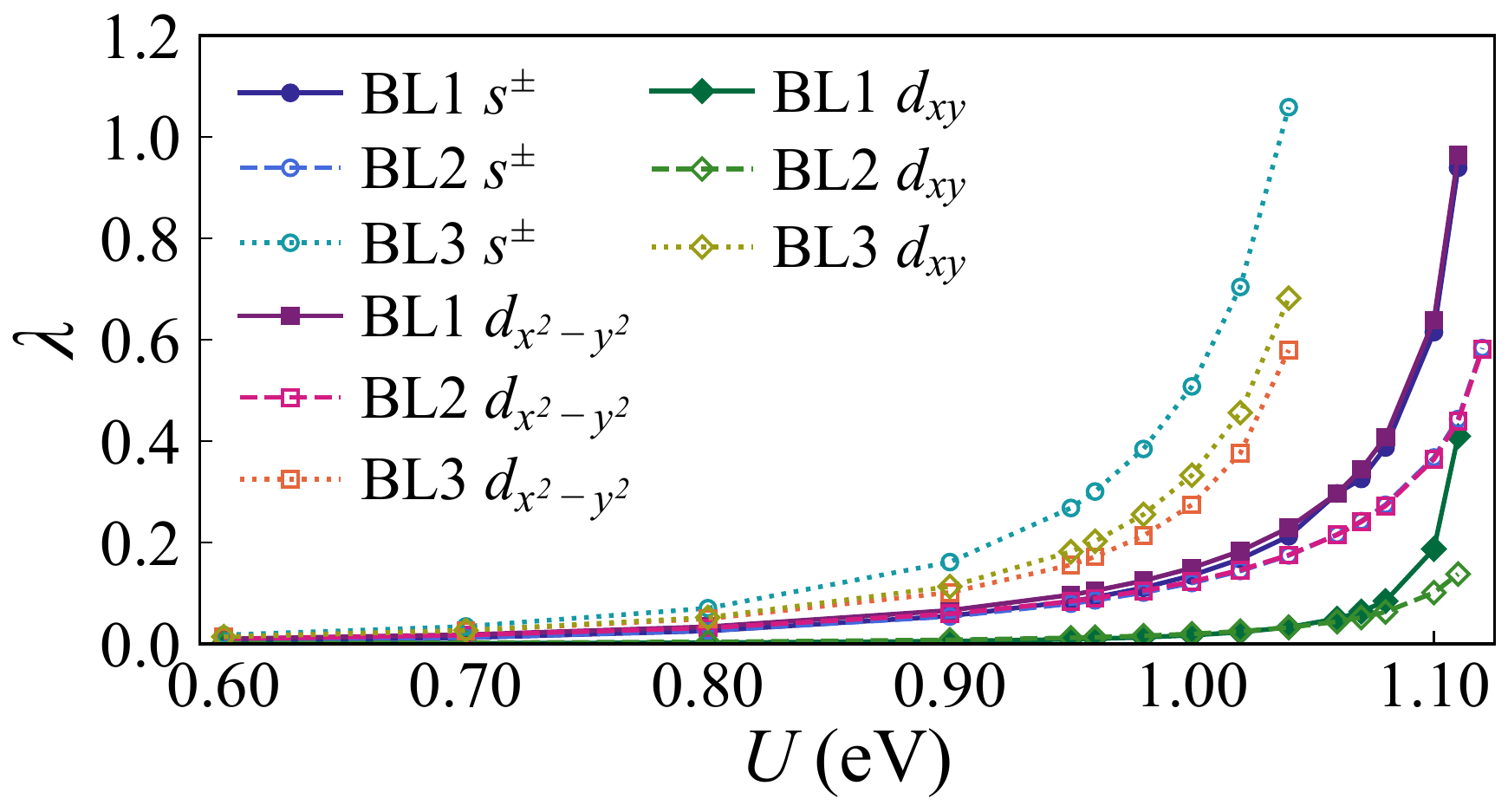}
\caption{Pairing eigenvalues $\lambda$ vs. the interaction strength $U$ for each bilayer model (BL1, BL2, and BL3). The $s^{\pm}$, $d_{x^2-y^2}$, and $d_{xy}$ pairing channels are shown in each case.}
\label{Scan_U_pairing}
\end{figure}

By solving the eigenvalue problem in Eq.~(\ref{eq:pp}) within the RPA framework for the BL models, we first calculated the pairing instabilities in different singlet pairing channels for each BL, as shown in Fig.~\ref{Scan_U_pairing}. 
For the surface BL3, the leading pairing instability occurs in the $s^{\pm}$ channel, with a well-separated subleading $d_{xy}$ state. For both the middle BL2 and interfacial BL1, the $d_{x^2-y^2}$ and $s^{\pm}$ pairing states are nearly degenerate with almost identical eigenvalues $\lambda$ even as $U$ increases, while the $d_{xy}$ state remains well separated from these two channels. The rapid enhancement of $\lambda$ with increasing $U$ reflects the spin-fluctuation-driven superconducting pairing in this system as the critical $U_c$ for the spin-density wave phase is approached. For BL2, the system becomes magnetically unstable for $U$ larger than $U_c = 1.12$~eV.

\begin{figure}
\centering
\includegraphics[width=0.42\textwidth]{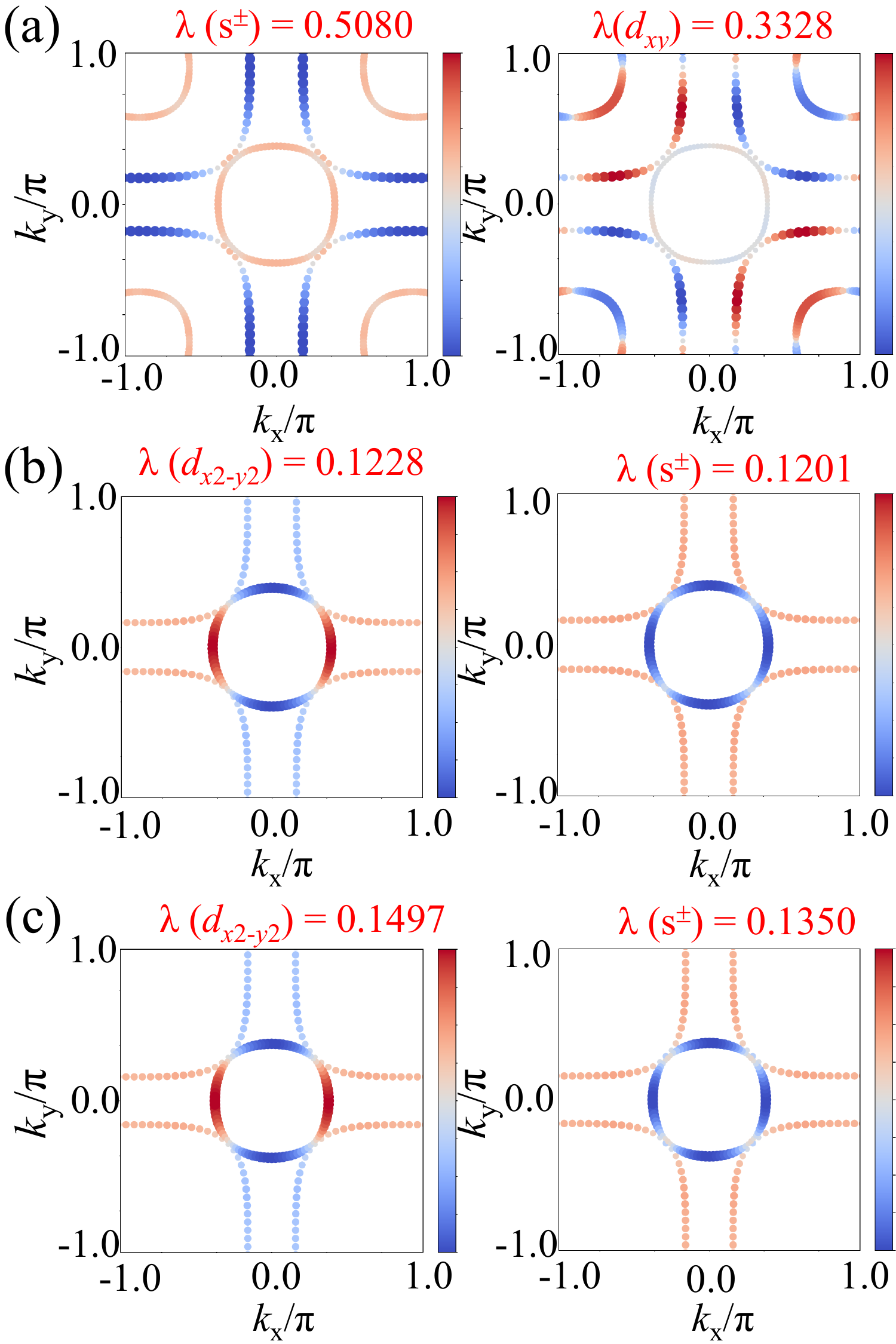}
\caption{The RPA-calculated leading and subleading superconducting singlet gap structures $g_\alpha(\mathbf{k})$ on the Fermi surfaces for (a) BL3, (b) BL2, and (c) BL1, respectively. Here, $U = 1.00$, $U'=U/2$, $J=J' = U/4$ are used at temperature $T=0.01$~eV, as in our previous studies with RPA. The meaning of the various couplings is standard and explained in the SM. The sign and magnitude of $g_\alpha({\bf k})$ is indicated by the color and its intensity  (red = positive, blue = negative).}
\label{RPA}
\end{figure}

To further understand the pairing characteristics, we also present the superconducting gap structures at $U = 1.0$ eV for each BL in Fig.~\ref{RPA}. As expected from the previous discussion, we find strong pairing in BL3, as shown in Fig.~\ref{RPA}(a), because the $\gamma$ pocket is present. Pairing is much weaker in BL1 and BL2 (see Figs.~\ref{RPA}(b-c)) because the $\gamma$ pocket is absent. For the surface BL3, the leading $s^{\pm}$ pairing state exhibits a sign change between the $\beta$ and $\gamma$ sheets, similar to that found in the high-pressure bulk phase~\cite{Zhang:nc24}. In addition, this leading $s^{\pm}$ state exhibits accidental nodes on the $\beta$ sheets. For BL3 the secondary pairing channel is $d_{xy}$, while for BL1 and BL2 it is $d_{x^2-y^2}$. The gap structures of the two strongly competing $d_{x^2-y^2}$ and $s^{\pm}$ pairing states in BL2 and BL1 are shown in Figs.~\ref{RPA}(b) and \ref{RPA}(c), respectively. For the $d_{x^2-y^2}$ pairing channel, the superconducting gap exhibits sign changes on the $\alpha$ and $\beta$ sheets, whereas the $s^{\pm}$ pairing state is characterized by a sign change between the $\alpha$ and $\beta$ sheets.

{\it DCA Results--} As discussed in Ref.~\cite{Maier:PRB2019}, incipient bands (close to, but not crossing, the Fermi level) may contribute to pairing if the states are within the dynamic range of the pairing interaction. Within our Fermi-surface-restricted RPA treatment, however, the incipient $\gamma$ bands in BL1 and BL2 do not explicitly enter in Eq.~(\ref{eq:pp}), even when the $\gamma$ band lies very close to the Fermi level, as in BL1. To study the effects of this incipient band on the leading pairing state, we therefore used Dynamical Cluster Approximation (DCA) quantum Monte Carlo \cite{Hettler:prb98,Maier:RMP,Maier:arxiv25,Maier:prb11,Troyer:prl2005,Liu:pre2016,Gull:prb2011,Gull:rmp2011,Jarrell:prb2001,Hahner:cpc2020} calculations for the BL two-orbital models. We discuss our DCA results for the pair-field susceptibility defined as 
\begin{equation}\label{eq:Palpha}
P_\alpha(T) = \int_0^\beta d\tau \langle {\cal T}_\tau \Delta^{\phantom\dagger}_\alpha(\tau)\Delta^{\dagger}_\alpha(0)\rangle,
\end{equation}
with the pair operator
\begin{equation}
\Delta^{\dagger}_\alpha = \frac{1}{\sqrt{N}}\sum_{(\bm k),\ell\ell'} g^{\ell\ell'}_\alpha(\bm k) c^\dagger_{\bm k \ell\uparrow}c^\dagger_{-\bm k \ell'\downarrow}\,.
\end{equation}
The momentum space Fourier representation with wave-vector ${\bm k}=(k_x, k_y, k_z)$ was used, with $k_z=0$ or $\pi$ representing bonding and anti-bonding combinations of the two layers in the BL, respectively, $c^\dagger_{{\bm k}\ell\sigma}=1/\sqrt{N}\sum_i c^\dagger_{i\ell\sigma}e^{i{\bm k}\cdot{\bm r}_i}$, and $g_\alpha^{\ell\ell'}({\bm k})$ is a pair form-factor for $s^\pm$, $d_{x^2-y^2}$, and $d_{xy}$ symmetries. We employ an eight-site ($N_{\mathrm{c}}=8$) tilted-square DCA cluster defined by the in-plane supercell vectors $(2,2)$ and $(2,-2)$ (see Fig.~72 of Ref.~\cite{Dagotto:rmp94}). The cluster is a $45^{\circ}$-rotated $\sqrt{8}\times\sqrt{8}$ square and contains eight in-plane unit cells. With four correlated $e_g$ orbitals per bilayer unit cell, the cluster thus contains 32 spatial orbital degrees of freedom.

\begin{figure}
\centering
\includegraphics[width=0.45\textwidth]{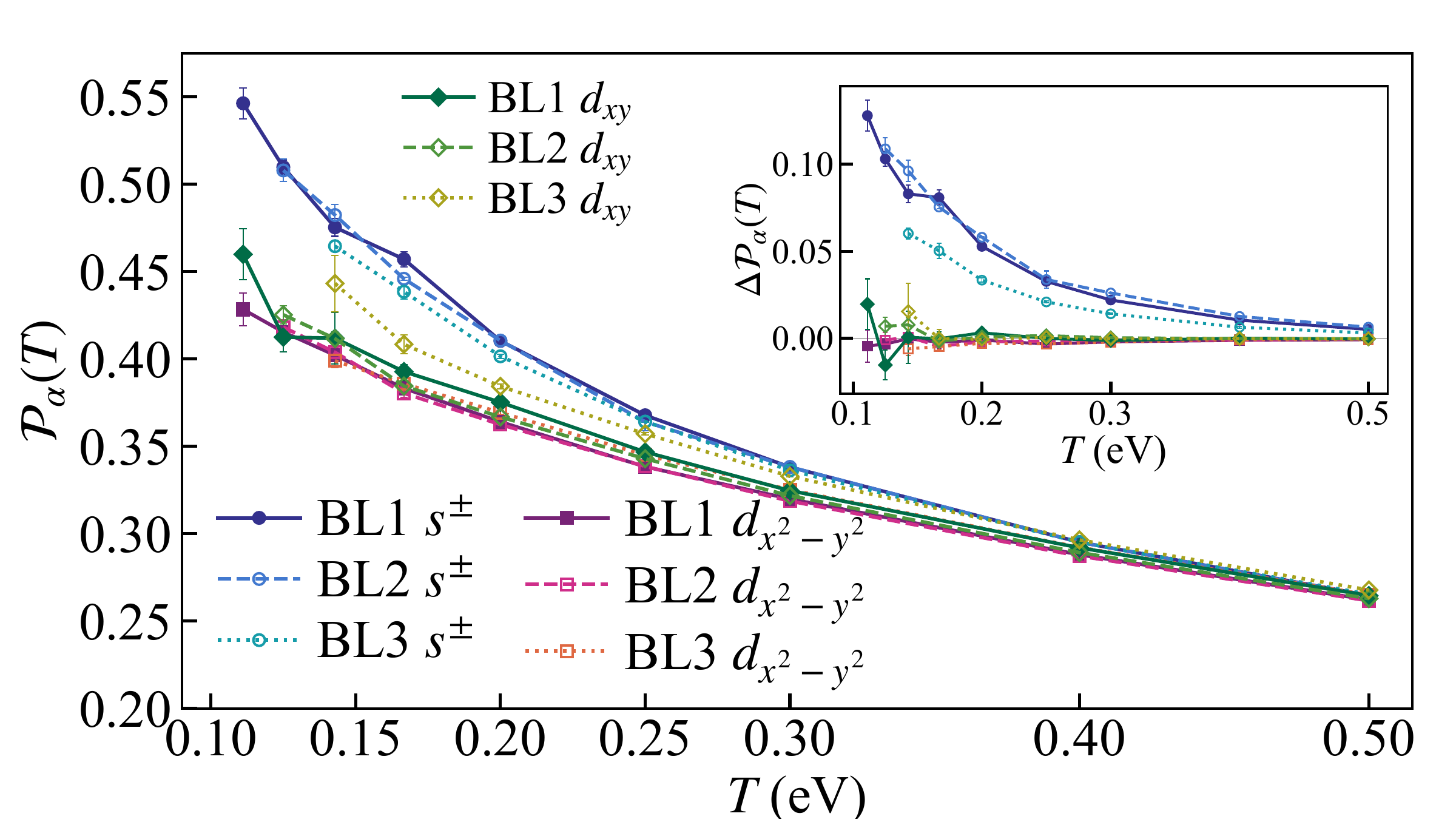}
\caption{Pair-field response of each BL. The main panel shows the static, uniform ($q = 0$, $\omega_m = 0$) lattice pair-field susceptibility $P_\alpha(T)$ for the $s^{\pm}$, $d_{x^2-y^2}$, and $d_{xy}$ pairing channels. {\it Inset:} the interaction-induced enhancement ($\Delta P_\alpha(T) = P_\alpha(T) - P_\alpha^{(0)}(T)$), where $P_\alpha^{(0)}(T)$ is the disconnected lattice-bubble contribution calculated from the dressed single-particle Green's functions. Error bars represent one standard error estimated using delete-one jackknife resampling of independent two-particle measurement replicas. The single-particle input is held fixed, and its uncertainty is therefore not included in the error bars. Lines are guides to the eye. Here, $U=3$ eV, $U'=2$ eV and $J=0.5$ eV are used for DCA calculations, as in previous studies~\cite{Maier:arxiv25,Sakakibara:prl24}.}
\label{DCA}
\end{figure}

Figure~\ref{DCA} shows the temperature dependence of the pair-field susceptibility $P_\alpha(T)$ for different pairing channels in each BL. For all three BLs, the pair-field susceptibilities increase as the temperature decreases, indicating enhanced superconducting pairing correlations at lower temperatures. The three pairing channels exhibit comparable susceptibilities at high temperatures. However, the $s^{\pm}$ susceptibility becomes clearly dominant upon cooling, suggesting a preference for $s^{\pm}$ pairing symmetry in all BLs, even in the absence of Fermi surface $\gamma$ pockets in BL1 and BL2, clearly indicating the importance of incipient bands beyond Fermi surface-restricted weak-coupling approximations.

To further examine the role of interactions in superconducting pairing, we calculated the interaction-induced enhancement of the pair-field susceptibility, $\Delta P_\alpha(T)=P_\alpha(T)-P_\alpha^{(0)}(T)$, as shown in the inset of Fig.~\ref{DCA}. A pronounced enhancement is observed in the $s^{\pm}$ channel for all three BLs, where the interaction-induced enhancement increases rapidly as the temperature decreases. By contrast, the $d_{x^2-y^2}$ and $d_{xy}$ channels exhibit weaker enhancements within the estimated uncertainties, remaining close to zero over most of the temperature range. These results indicate that the $s^{\pm}$ pairing channel is preferentially enhanced by electronic interactions, supporting the dominant role of $s^{\pm}$ pairing correlations in all three BLs.

{\it Conclusion--} 
In summary, we systematically studied LNO thin films grown on a real LSAO substrate. The low-energy electronic states near the Fermi level are dominated by the Ni-$e_g$ orbitals, while the states associated with the LSAO substrate lie {\it at much lower energies}. This casts doubt on the hypothesis that charge migration from LSAO could induce partial doping of the nickelate thin film. Remarkably, the Fermi surface topology of the BL nickelate thin film still closely resembles that of the high-pressure bulk phase, featuring similar $\alpha$ and $\beta$ sheets, together with a $\gamma$ pocket. This hole-like $\gamma$ pocket around the M point {\it originates from the surface BL3 portion of the thin film}, and is absent in the other BLs. The reason for this behavior is a (small but important) charge disproportionation among the different BLs, with hole doping in the surface BL3 and electron doping in the middle BL2 and interfacial BL1. This charge rearrangement is possible due to the lack of inversion symmetry in the thin film between top and bottom bilayers. Our weak-coupling RPA calculations reveal a strong $s^{\pm}$ superconducting pairing instability in the surface BL3, while weaker and nearly degenerate $d_{x^2-y^2}$ and $s^{\pm}$ pairing states are found in the middle BL2 and interfacial BL1. In contrast, our non-perturbative DCA calculations unveil robust $s^{\pm}$ pairing in all three layers, despite the fact that the $\gamma$ band does not cross the Fermi level in BL1 and BL2. To conclude, {\it (i)} the LSAO substrate does not transfer charge to LNO, and {\it (ii)} the $s^{\pm}$-wave pairing is the leading pairing state found in the thin film.

{\it Acknowledgment--}
This work was supported by the U.S. Department of Energy, Office of Science, Basic Energy Sciences, Materials Sciences and Engineering Division.  This manuscript has been authored by UT-Battelle, LLC, under contract DE-AC05-00OR22725 with the US Department of Energy (DOE). An award of computer time was provided by the INCITE program. This research also used resources of the Oak Ridge Leadership Computing Facility, which is a DOE Office of Science User Facility supported under Contract DE-AC05-00OR22725. The US government retains, and the publisher, by accepting the article for publication, acknowledges that the US government retains a nonexclusive, paid-up, irrevocable, worldwide license to publish or reproduce the published form of this manuscript, or allow others to do so, for US government purposes. DOE will provide public access to these results of federally sponsored research in accordance with the DOE Public Access Plan (https://www.energy.gov/DOE Public Access Plan).

{\it Data availability--}
The dataset of the main findings of this study will be openly available in the Zenodo Repository upon acceptance of the final version of the paper for publication. Until that time, the data will be made available upon reasonable request. In addition, the hopping and crystal field parameters for this work are available in the Supplementary Material and Zenodo Repository.

\end{document}